\documentclass[aps,prd,twocolumn,preprintnumbers,amsmath,amssymb,superscriptaddress,groupedaddress,floatfix,nofootinbib]{revtex4}

\usepackage[normalem]{ulem}
\usepackage[utf8]{inputenc}
\usepackage{epsfig,psfrag,graphics,verbatim}
\usepackage{graphicx}
\usepackage[T1]{fontenc}
\usepackage{subfigure}
\usepackage{dcolumn}
\usepackage{bm}
\usepackage{xcolor}
\usepackage{float}
\usepackage{placeins}
\usepackage{booktabs}
\usepackage{adjustbox}
\usepackage{amsmath}
\usepackage{subcaption}
\usepackage{makecell}
\usepackage{eucal}
\usepackage{varioref}
\usepackage{hyperref}
\usepackage{cleveref}
\usepackage{soul}

\newcommand{\nn}{\nonumber \\}

\date{June 2025} 

\begin{document}
\title{Probing the $H_0$ Tension with Holographic Dark Energy in Unimodular Gravity: Insights from DESI DR2}

\author{Francisco Plaza}
\email{Fran22@fcaglp.unlp.edu.ar}
\affiliation{%
 Facultad de Ciencias Astronómicas y Geofísicas, Universidad Nacional de La Plata, Observatorio Astronómico, Paseo del Bosque,\\
B1900FWA La Plata, Argentina \\
}%
\affiliation{Consejo Nacional de Investigaciones Científicas y Técnicas (CONICET), Godoy Cruz 2290, 1425, Ciudad Autónoma de Buenos Aires, Argentina}
\author{Gabriel Le\'on}
\affiliation{%
 Facultad de Ciencias Astronómicas y Geofísicas, Universidad Nacional de La Plata, Observatorio Astronómico, Paseo del Bosque,\\
B1900FWA La Plata, Argentina \\
}%
\affiliation{Consejo Nacional de Investigaciones Científicas y Técnicas (CONICET), Godoy Cruz 2290, 1425, Ciudad Autónoma de Buenos Aires, Argentina}

\author{Lucila Kraiselburd}
\affiliation{%
 Facultad de Ciencias Astronómicas y Geofísicas, Universidad Nacional de La Plata, Observatorio Astronómico, Paseo del Bosque,\\
B1900FWA La Plata, Argentina \\
}%
\affiliation{Consejo Nacional de Investigaciones Científicas y Técnicas (CONICET), Godoy Cruz 2290, 1425, Ciudad Autónoma de Buenos Aires, Argentina}

\begin{abstract}
Motivated by the recent baryon acoustic oscillation measurements of DESI DR2 collaboration, this works presents an extended analysis of a cosmological model based on holographic dark energy within the framework of  Unimodular Gravity. We probe the model with an extensive set of observations: cosmic chronometers, Pantheon Plus+SH0ES Type Ia supernovae, DESI DR2 BAO distances, quasar X-ray/UV fluxes (two independent calibrations), and \textit{Planck} 2018 CMB data. The results are analyzed to assess the model’s ability to alleviate the Hubble tension and, in comparison with the standard $\Lambda$CDM framework, to determine which of the two scenarios is preferred according to Bayesian evidence. We conclude that the present implementation of holographic dark energy in Unimodular Gravity, while theoretically appealing, does not alleviate the Hubble tension and is not statistically preferred by Bayesian criteria when compared with the standard $\Lambda$CDM model. Nevertheless, in neither case does the preference become very strong or conclusive against it.

\end{abstract}
\maketitle
\section{Introduction}
\label{secI}
One of the most important open problems in contemporary cosmology is to explain the physics underlying the accelerated expansion of the Universe, discovered in 1998 \cite{rii1998AJ....116.1009R,Perlmutter1999ApJ...517..565P}. Following this discovery, the standard cosmological model incorporated a cosmological constant into the Einstein--Hilbert action, thereby achieving remarkable agreement with observations and establishing the $\Lambda$CDM cosmological model. In this scenario, late--time acceleration is reproduced with only a few free parameters, which has made the model the prevailing paradigm.
However, neither the mechanism driving the acceleration nor the nature of the cosmological constant are understood. Moreover, in recent years, the $H_0$ tension has emerged, referring to the discrepancy greater than 4$\sigma$ between the Hubble parameter estimate reported by \citep{3riess2024IAUS..376...15R} and other model‐dependent estimates\footnote{Specifically, the $H_0$ estimate obtained within the $\Lambda$CDM framework that displays the largest tension comes from the \textit{Planck} Collaboration \cite{Planckcosmo2018}, which is based on CMB observations.}. Then, a significant number of proposals have been considered to address these issues, which can be grouped into three types of models: (i) models that introduce new components in the energy-momentum tensor, such as scalar fields collectively referred to as dark energy; (ii) theories that attribute the phenomenon to modified spacetime dynamics, thus requiring an extension of General Relativity; and (iii) scenarios that identify the cosmological constant with the vacuum energy of quantum fields. The last category yields the most inaccurate prediction in the history of physics \cite{Weinberg89,Bahcall1999,1995ostrikerNatur.377..600O}. Indeed, no satisfactory mechanism exists such that vacuum energy can drive cosmic acceleration without encountering severe theoretical difficulties. Models of this kind have been proposed, e.g. \cite{2017unrPhRvD..95j3504W}, but they have revealed internal inconsistencies \cite{2020vacumEPJC...80...18B}.

Dark energy (DE) models, on the other hand, postulate additional degrees of freedom that must account for roughly $70\%$ of the present energy budget. This raises new questions regarding the particle nature of these fields and the reason for their non-detection to date. Still, the regained interest in DE models stems from recent DESI collaboration work \citep{DR22025arXiv250314738D}, which presents a new set of baryon acoustic oscillation (BAO) measurements together with a detailed statistical analysis. That study performed a Bayesian comparison between a dynamical DE scenario, characterized by a time-dependent equation of state, and the standard $\Lambda$CDM model. The principal finding is a statistical preference for the aforementioned alternative model over the standard cosmological framework, which becomes very strong in certain cases according to the  Deviance Information Criterion $\Delta\mathrm{DIC}$. This result is notable because, even though the alternative model introduces additional free parameters, the $\Lambda$CDM model is nevertheless disfavored.

The remaining class of proposals, which is the focus of this work, has been extensively studied in the literature. These models attribute the accelerated expansion to unknown features of spacetime itself, which can be described by extensions of General Relativity (GR).  Examples include $f(R)$ theories \cite{husawicki2007PhRvD..76f4004H,2007JETPL..86..157S,Sotiriou2010,DeFelice2010}, scalar-vector-tensor (SVT) theories \cite{2006STVGJCAP...03..004M,Heisenberg2018_SVT,KaseTsujikawa2018}, and Unimodular Gravity \cite{UG1,UG2,UG3,UG4,Finkelstein1971}, the latter of which will be introduced and analyzed in this work. These alternative gravity theories are inspired by the idea that GR is perhaps not the ultimate theory of gravity given the well-known challenges in developing a consistent quantum description.

It is important to emphasize that the DESI result, which shows a statistical preference for the dynamical DE scenario, does not imply that this class of models is necessarily correct \cite{Lee:2022cyh}. Instead, it suggests that any framework with comparable behavior and predictions is similarly favored, including alternative gravity models that exhibit analogous dynamics. For instance, Plaza and Kraiselburd \citep{Plaza:2025gcv} recently reported a very strong preference for a cosmological model based on $f(R)$ gravity over the $\Lambda$CDM paradigm.

In this work, we will employ the Unimodular Gravity (UG) framework. This theory was originally proposed by A.~Einstein himself in 1919 \cite{UG0} and is arguably the most conservative modification of GR. In its modern formulation, UG can be obtained from the Einstein--Hilbert action by restricting variations to those that preserve the volume element. Thus, UG is invariant only under volume-preserving diffeomorphisms, and consequently, it possesses fewer symmetries than GR.  Due to the structure of the theory, UG permits a possible non-conservation of the canonical energy-momentum tensor, typically described by a \textit{diffusion term}. If one adopts energy-momentum conservation as an additional postulate, the standard Einstein field equations with a cosmological constant are recovered, but the latter is interpreted as an integration constant \cite{UG4,Weinberg1989,ellis2010,ellis2015} (i.e. not sourced by the energy-momentum of vacuum flucutations). Consequently, all successful predictions of GR are retained within UG. Conversely, removing the conservation assumption leads to the non-conservative formulation of UG, in which the diffusion terms acts as a ``dynamical cosmological constant.'' The implementation of UG in cosmology and its connection with the cosmological constant have been recognized for decades \cite{Weinberg89,ellis2010,UG4,Smolin2009,Uzan2010}; recently, the topic has been rediscovered and is receiving significant attention \cite{ellis2015,Corral20,Nucamendi20,sudarskyPRL1,sudarskyPRL2,sudarskyH0,aperez2021,leonBengochea_2023,landau2022cosmological,Leon_2022,pia2023,Fabris:2021atr,Fabris:2025jhu,deCesare:2021wmk,deCesare:2023fbq, Malekpour:2023lsf,Albertini:2024yfm,Chakraborty:2024vqa, divalentinoUG, velten2024}.

One particular implementation of non-conservative UG in cosmology is the model introduced in Ref. \cite{velten2024}. This model interprets the diffusion term as an energy exchange between matter and a particular form of DE called holographic dark energy. The latter is based on the holographic principle of 't Hooft and Susskind \cite{Susskind:1994vu,tHooft:1993dmi}, which generalizes the Hawking–Bekenstein entropy $S \sim A/(4G)$~\cite{Hawking:1971tu,Bekenstein:1973ur,Bekenstein:1974ax} by imposing that the maximal entropy of a region of spacetime scales with its boundary area. This prevents quantum field theory from overcounting UV degrees of freedom and ensures that no region can contain more energy than a black hole of the same size. The cut-off scale proposed in \cite{velten2024} is the Ricci scalar. In our view, this is a natural choice within the UG framework because $R$ enters directly in the diffusion term responsible for the non-conservative dynamics of the energy-momentum tensor.

In Ref. \cite{velten2024}, the model was tested using only cosmic chronometer data. Here, we perform a comprehensive analysis that includes additional late‐time observations such as: Type Ia supernovae, BAO, and active galactic nuclei (AGN) measurements; moreover, we also include early‐time data from the cosmic microwave background (CMB), providing a far more stringent evaluation of the model. We further extend the model to distinguish between baryonic and cold dark matter and to account for radiation in the total energy density budget, a fact that had not been studied until now. An objective of the present work is to analyze whether this UG-based holographic DE model is statistically favored when including the DESI DR2 dataset \citep{DR22025arXiv250314738D}. The motivation is because, as explained above, this dataset seems to favor dynamical DE models. Another goal of our work is to explore the capability of the model to alleviate the well‐known Hubble tension. 

For this purpose, we perform Markov Chain Monte Carlo (MCMC) parameter estimations for both the $\Lambda$CDM and the UG model, employing the previously mentioned observational data. We subsequently evaluate whether either model alleviates the Hubble tension, focusing on fits that incorporate CMB data, and conduct a Bayesian model comparison to determine which framework is more strongly supported by the observations. It is also worth noting that the present analysis makes use of two distinct methodologies to incorporate AGN data and we explore how the choice of method affects the results.

The article is organized as follows: Section \ref{secII} introduces the cosmological framework of non-conservative UG, including a dedicated subsection that embeds holographic DE in this setting. Section \ref{secIII} describes the observational data employed, with four subsections addressing Cosmic Chronometers, Type Ia Supernovae, BAO, AGN measurements and CMB data. Section \ref{secIV} presents the method and MCMC analysis leading the resulting constraints, with particular attention to the $H_0$ tension. Finally, Section \ref{secV} summarizes our findings and present the main conclusions.

\section{Cosmological model: non-conservative unimodular gravity }
\label{secII}

In this section, we introduce our dark energy model within the non-conservative UG scenario.

The UG gravitational equations are given explicitly by:
\begin{equation}\label{UGefe}
	R_{\mu\nu} - \frac{1}{4}g_{\mu\nu}R = \kappa \left( T_{\mu\nu} - \frac{1}{4} g_{\mu\nu}T \right),
\end{equation}
where $\kappa \equiv 8 \pi  G$; $R_{\mu \nu}$ and $R$ represent the Ricci tensor and Ricci scalar respectively;  $T_{\mu \nu}$ and $T$ denote the energy–momentum tensor of matter fields and its trace respectively. An important characteristic of UG is the possible non-conservation of the canonical energy–momentum tensor, expressed as:
\begin{equation}\label{conservEM}
	\nabla^\mu T_{\mu\nu} = \nabla_\nu Q,
\end{equation}
where $Q(x)$ represents a scalar diffusion term whose nature and origin could be related to quantum gravitational effects (e.g. fundamental spacetime discreteness at Planck scales \cite{sudarskyPRL1,sudarskyPRL2,aperez2021,Bengochea:2024msf}). If $Q$ is a  constant, traditional GR is recovered, and the standard conservation law holds. In other words, one can impose the conservation of $T_{\mu \nu}$ but it would be an extra assumption that is not required by the theory. 

The UG equations can be rewritten in such a way that the term $Q$ appears explicitly \cite{UG3,UG4,Finkelstein1971,GL2023} this is,
\begin{equation}\label{UGlambdax}
		R_{\mu\nu} - \frac{1}{2}g_{\mu\nu}R + \kappa Q g_{\mu \nu} = \kappa T_{\mu\nu}.   
\end{equation}
Taking the trace of Eq. \eqref{UGlambdax}, one finds that 
\begin{equation}\label{defQ}
   Q(x) = ( T + \kappa^{-1} R)/4
\end{equation}
Replacing $Q(x)$ back into Eq. \eqref{UGlambdax}, one recovers the original trace-free UG equations \eqref{UGefe}. The utility of the equivalent UG equations \eqref{UGlambdax} is that they resemble to the Einstein's Field Equations but with a ``dynamical cosmological constant'' contained in the term $Q(x)$ (which also allows for the violation of the energy-momentum conservation).

Within a cosmological framework, we assume a spatially flat Friedmann--Lemaître--Robertson--Walker (FLRW) metric for the background spacetime:
\begin{equation}\label{FLRW}
	ds^2 = -dt^2 + a^2(t)(dx^2 + dy^2 + dz^2).
\end{equation}

The total matter content in the universe is modeled by hydrodynamical matter, with the energy-momentum tensor given by
\begin{equation}\label{EMtensor}
	T_{\mu \nu} = (\rho + p ) u_\mu u_\nu + p g_{\mu \nu} + \pi_{\mu \nu},
\end{equation}
where $\rho$ is the total energy density, $p$ is the total pressure in the rest frame of the fluid, and $\pi_{\mu \nu}$ is the total anisotropic stress
\begin{equation}
	\rho = \sum_I \rho_I, \qquad
	p = \sum_I p_I, \qquad
    \pi_{\mu \nu} = \sum_I \pi^I_{\mu \nu}. 
\end{equation}
so each matter species is labeled by the index $I$;  $u^\mu$ is the 4-velocity of total matter fluid (normalized as $u^\mu u_\mu =-1$) relative to the observer. For each component, we consider a barotropic equation of state $p_I = \omega_I \rho_I$, with $\omega_I$ a constant. 

At the background level, we assume that  both, the total energy-momentum tensor and the diffusion term $Q$, are spatially isotropic and homogeneous; so the 4-velocities of each matter species are equal, and their anisotropic stress vanishes.  The resulting Friedmann equations become
\begin{equation}\label{F1}
	3H^2 = \kappa (\rho + Q),
\end{equation}
\begin{equation}\label{F2}
	2 \dot H + 3 H^2 = \kappa (-p + Q),
\end{equation}
The Hubble parameter is defined as $H \equiv \dot a/a$, the dot denotes derivatives with respect to cosmic time $t$. We can combine Eqs. \eqref{F1} and \eqref{F2} to obtain
\begin{equation}\label{F3}
	\dot H = -\frac{\kappa}{2} (\rho + p)
\end{equation} 
We note that in \eqref{F3}, the term $Q$ is absent; in fact, the equation has the same form as in standard GR with a cosmological constant. Therefore, one may interpret $Q$ as an additional energy density component  (so $\rho_Q \equiv Q>0$) characterized by the equation of state $\rho_Q = -p_Q$, i.e.\ $\omega_Q = -1$. This property distinguishes $Q$ from conventional dynamical dark energy parametrizations. Whereas the latter typically adopt the Chevallier-Polarski-Linder (CPL) parametrization \cite{Chevallier:2000qy,Linder:2002et,DESI:2024mwx,Giare:2025pzu} $\omega(a) = \omega_0 + \omega_a(1 - a)$, here one finds $\omega_Q = -1$ exactly, even though the term $Q$ acts effectively as a ``dynamical cosmological constant.''

From the non-conservation of the energy-momentum tensor, Eq. \eqref{conservEM}, we derive
\begin{equation}\label{rhomov}
	\dot \rho + \dot Q + 3 H(\rho + p) = 0.
\end{equation}

In order to close the system of equations, we need to specify $Q(t)$. In the next subsection, we characterize this term based on the model of Ref. \cite{velten2024}.

\subsection{Holographic dark energy in Unimodular Gravity}

The model analyzed in Ref. \cite{velten2024} associates the $Q$ term with an interacting energy flow between matter and an additional component: the holographic dark energy, $\rho_H$. This form of dark energy is based on the holographic principle originally proposed by 't Hooft and Susskind \cite{Susskind:1994vu,tHooft:1993dmi}, which asserts that the maximum entropy within a volume is proportional to the area $A$ of its boundary. This principle is a generalization of the Hawking--Bekenstein entropy of a black hole, $S \sim A/(4G)$ \cite{Hawking:1971tu,Bekenstein:1973ur,Bekenstein:1974ax}. Hence, the holographic principle suggests that a naive application of QFT overcounts degrees of freedom at UV scales, and that a fundamental theory of quantum gravity must enforce an entropy (or energy) cutoff that prevents a region containing more energy than a black hole of the same size. In particular, Cohen, Kaplan, and Nelson argued that the total vacuum energy in a region of size $L^3$ should not exceed the mass of a black hole of the same size \cite{Cohen:1998zx}:
\begin{equation}
	E_{\rm vac} = \rho_H L^3 \leq M_{\rm BH} \simeq M_P^2 L,
\end{equation}
which yields an upper bound on the vacuum energy density $\rho_H$,
\begin{equation}\label{rhoHDE}
	\rho_H = 3c^2 M_P^2 L^{-2}
\end{equation}
where the number $3$ is introduced for convenience and $c^2$ represents a dimensionless parameter of order unity. Equation \eqref{rhoHDE} defines the holographic dark energy density for a given choice of length $L$ corresponding to an IR cutoff. The challenge is to identify the correct $L$ for cosmology\footnote{Notably, if one chooses $L$ to be on the order of the current Hubble radius $H_0^{-1} \simeq 10^{61} M_P^{-1}$, the energy density given by Eq. \eqref{rhoHDE} is of the same order as the cosmological constant today $\Lambda \sim 10^{-122} M_P^4 $.}, and various models of holographic dark energy in the context of GR have been studied \cite{Li:2004rb,delCampo:2011jp,Wang:2016och,Nojiri:2017opc,Fu:2011ab,Li:2024qus,Dai:2020rfo,Colgain:2021beg}. 

One particular choice is $R = L^{-2}$, with $R$ the Ricci scalar, which leads to the holographic Ricci dark energy in GR \cite{Gao:2007ep}. In Ref. \cite{velten2024}, the holographic Ricci dark energy model was implemented in the framework of UG through an interaction between $\rho_H$ and matter. In particular, this interacting model is expressed using the following continuity equations
\begin{subequations}\label{veltenecs}
 \begin{equation}
 	\frac{2}{3} \dot \rho_H = -\beta H \rho_{m},
 \end{equation}
 \begin{equation}
 	 \dot \rho_m + H \rho_m (4-\beta) = 0.
 \end{equation}
\end{subequations}
The resulting solutions for the evolution of $\rho_H$ and $\rho_m$  are
\begin{subequations}\label{veltensols}
	\begin{equation}\label{rhomsol}
	\rho_m (a) =  3 \kappa^{-1} H_0^2 \Omega_{m}  \bigg(\frac{a_0}{a}\bigg)^{(4-\beta)},
	\end{equation}
\begin{equation}\label{rhoHsol}
	\rho_H(a) =  3 \kappa^{-1} H_0^2  \left\{  \Omega_{H} + \frac{3 \beta \Omega_{m}}{2 (4-\beta) } \left[\bigg(\frac{a_0}{a}\bigg)^{(4-\beta)} - 1 \right] \right\},
	\end{equation}	
\end{subequations}
where $H_0$ and $\Omega_{m}$ are the present day values of the Hubble factor and the matter (pressure-less) density parameter respectively; $\Omega_{H}  \equiv \kappa \rho_{0H}/3 H_0^2$ is a constant and is related to the holographic dark energy today $\rho_{0H}$.  The dimensionless parameter $\beta>0$ characterizes the interaction between $\rho_H$ and $\rho_m$. 

We aim to find explicitly the term $Q$ that yields the solutions \eqref{veltensols}. We begin by combining Eqs.~\eqref{veltenecs} to obtain
\begin{equation}\label{rhomQ}
	\dot\rho_m + 3H \rho_m = -\frac{2}{3}\dot\rho_H - H \rho_m.
\end{equation}
Comparing Eq. \eqref{rhomQ} with Eq. \eqref{rhomov} for pressureless dust, we identify $\dot Q = (2/3) \dot\rho_H + H\rho_m$ or, expressed in terms of the scale factor $a$,
\begin{equation}\label{dQda}
	\frac{dQ}{da} = \frac{2}{3}\frac{d\rho_H}{da} + \frac{ \rho_m}{a}.
\end{equation}
Substituting solutions \eqref{veltensols} in the previous equation and integrating we find
\begin{equation}
	Q (a) = Q_0 - 3 \kappa^{-1} H_0^2 \Omega_{m} \left( \frac{\beta-1}{\beta-4} \right) \bigg(\frac{a_0}{a}\bigg)^{(4-\beta)},
	\end{equation}
where $Q_0$ is an integration constant that we can identify directly with the value of dark energy density today $Q(a_0) = 3 \kappa^{-1} H_0^2 \Omega_{0 \Lambda}$, therefore
\begin{eqnarray}\label{Qfinal}
	Q(a) &=& 3 \kappa^{-1} H_0^2 \left[  \Omega_{0 \Lambda} + \Omega_{m} \left( \frac{\beta-1}{\beta-4} \right)    \right] \nn
	&-& 3 \kappa^{-1} H_0^2 \Omega_{m} \left( \frac{\beta-1}{\beta-4} \right) \bigg(\frac{a_0}{a}\bigg)^{(4-\beta)}. 
\end{eqnarray}

With the diffusion term $Q$ at hand, we can reformulate the cosmological model of Ref. \cite{velten2024}. In particular, the interacting model based on holographic dark energy is equivalent to a cosmological model in UG in which the diffusion term $Q$, given by Eq.~\eqref{Qfinal}, couples exclusively to pressureless matter via
\begin{equation}
	\dot{\rho}_m + 3H\rho_m + \dot{Q} = 0.
\end{equation}
The solution to this equation, expressed in terms of the scale factor, is provided in Eq. \eqref{rhomsol}. The evolution equation for the holographic dark energy density $\rho_H$ can then be derived from Eq. \eqref{dQda}, yielding the solution in Eq. \eqref{rhoHsol} with $\Omega_{H} \equiv 2 \Omega_{\Lambda} + \Omega_{m}/2$. 

In the present paper, we extend the UG  model to include radiation in the simplest manner. Specifically, we assume that the radiation fluid does not couple to $Q$; hence, radiation follows the standard continuity equation $\dot \rho_r + 4H \rho_r = 0$, yielding
\begin{equation}
  \rho_r (a) =  3 \kappa^{-1} H_0^2 \Omega_{r}  \bigg(\frac{a_0}{a}\bigg)^{4},
\end{equation}
where $\Omega_{r}$ corresponds to the current value of the radiation density parameter. 

On the other hand, the evolution equation of matter, Eq. \eqref{rhomsol} includes barionic and cold dark matter, $\rho_m = \rho_b + \rho_{cdm}$, with evolution
\begin{subequations}\label{rhomsol2}
	\begin{equation}\label{rhomsolb}
		\rho_{b} (a) =  3 \kappa^{-1} H_0^2 \Omega_{b}  \bigg(\frac{a_0}{a}\bigg)^{(4-\beta)},
	\end{equation}
	\begin{equation}\label{rhomsolcdm}
		\rho_{cdm} (a) =  3 \kappa^{-1} H_0^2 \Omega_{cdm}  \bigg(\frac{a_0}{a}\bigg)^{(4-\beta)},
	\end{equation}
\end{subequations}	
where $\Omega_{b}$,  $\Omega_{cdm}$ are the present day values of the density parameter corresponding to baryonic and cold dark matter respectively. The total energy density is then $\rho = \rho_b + \rho_{cdm} + \rho_r$.  Thus, the complete model is described by 
\begin{equation}
	H(a)^2 = \frac{\kappa}{3} \left[ \rho_r (a) + \rho_b(a) + \rho_{cdm} (a) + Q(a) \right]
\end{equation}
with the diffusion term $Q$ as in Eq.~\eqref{Qfinal}, which incorporates the present dark energy density and $ \Omega_{m} =  \Omega_{b} + \Omega_{cdm}  $. It is also worth mentioning that if $\beta =1$ then $Q$ is a constant (equal to the present day value of the cosmological constant), see Eq. \eqref{Qfinal}. Additionally, if $\beta =1$ Eqs. \eqref{rhomsol2} evolve in the usual manner. Thus, for $\beta = 1$ the UG cosmological model is exactly equivalent to the standard $\Lambda$CDM model.

At the level of perturbations, the term $Q$ in Eq.~\eqref{Qfinal} does not affect scalar or tensor perturbations because $Q$ is homogeneous and isotropic, implying $\delta Q = 0$. However, this conclusion is based on a background-level analysis. If, on the other hand, we rewrite Eq.~\eqref{Qfinal}, using the definition of $Q_0$ and Eq.~\eqref{rhomsol}, as
\begin{equation}\label{Qrhom}
	Q = Q_0 + \left( \frac{\beta-1}{\beta-4} \right) \rho_{m},
\end{equation}
then, in principle, we find that 
\begin{equation}\label{deltaQ}
		\delta Q =  \left( \frac{\beta-1}{\beta-4} \right) \delta \rho_{m},
\end{equation}
which could certainly modify the evolution equations for the perturbations.

At this point, it is worth mentioning that a similar model, based also on UG, has been analyzed in Ref. \cite{divalentinoUG}. Specifically, in that work the diffusion term $Q$ takes the form $Q = Q^* + \epsilon \rho_{\mathrm{cdm}}$,
where $Q^*$ is an integration constant and $\epsilon$ is a free dimensionless parameter constrained by observations. Comparing this expression with Eq. \eqref{Qrhom}, we note that both are linear in the matter energy density. However, our model includes baryonic and cold dark matter, whereas the model in \cite{divalentinoUG} considers dark matter only. Also, the model in Ref.~\cite{divalentinoUG} is motivated phenomenologically and does not mention a connection with the holographic Ricci dark energy model. Nevertheless, their analysis includes the modified evolution equations for scalar perturbations due to the term $\delta Q = \epsilon\,\delta \rho_{\mathrm{cdm}}$; using recent data, the observational constraint on the model parameter is of order $\epsilon\sim10^{-4}$. Given the similar form of $Q$ between our model and that of Ref. \cite{divalentinoUG}, we therefore expect the parameter of our model to be very close to unity, or equivalently $\beta - 1 \ll 1$, see Eq. \eqref{deltaQ}. Moreover, since $\delta \rho_{m}\ll\rho_{m}^{(0)}$, with $\rho_{m}^{(0)}$ denoting the background matter energy density, then $(\beta-1) \delta \rho_{m} \ll (\beta-1) \rho_{cdm}^{(0)}$, this is, $(\beta-1) \delta \rho_{m}$ can be regarded at the same footing as  second-order perturbations. Consequently, we consider that a statistical analysis at the background level in $Q$ provides a good approximation. In particular, for the present paper, we proceed with the expression for $Q$ given in Eq.~\eqref{Qfinal}, which is proportional to $(\beta-1) \rho_{m}^{(0)}$ and is therefore homogeneous and isotropic. Thus, the evolution equations for the scalar and tensor perturbations remain unchanged.

Finally, another aspect of UG that deserves discussion, particularly in the context of cosmological perturbation theory, is the gauge choice and the so-called ``unimodular constraint.'' The latter refers to a condition in UG expressed as 
\begin{equation}\label{UGconstraint}
\varepsilon_{abcd} = \epsilon_{abcd}^{(g)},
\end{equation}
where $\varepsilon_{abcd}$ is a fiducial 4-volume element supplied by the theory, and $\epsilon_{abcd}^{(g)}$ is the 4-volume element associated with the spacetime metric $g_{ab}$. The 4-volume $\varepsilon_{abcd}$ is fixed (i.e., non-dynamical in the sense of a gravitational theory). In the literature \cite{UG1,UG2,UG3,UG4,Finkelstein1971},  the UG constraint \eqref{UGconstraint} is often written as $ \sqrt{-g} = f$, with $f$ a fixed function and $g$ the determinant of $g_{ab}$ in a chosen coordinate system. Consequently, within UG cosmological perturbation theory it is frequently argued that only gauge choices satisfying $\delta\sqrt{-g} = 0$ are admissible \cite{UGbranden,UGindios}. However, in Ref. \cite{GL2023}, it is shown that this latter claim is incorrect, and that the unimodular condition \eqref{UGconstraint} imposes no restrictions on the choice of gauge for perturbation analysis within UG. Furthermore, in practical calculations, one can always proceed as if, from the outset, the theory had been defined with a fiducial volume form $\varepsilon_{abcd}$ that ``happened to coincide'' with the one required to satisfy the constraint equation \eqref{UGconstraint} \cite{GL2023}.

\section{Observational Data}
\label{secIII}
This section details the observational data of different nature used to estimate both the cosmological parameters and the free parameter of the theory. There is data from the early universe as well as the late universe.
\subsection{Cosmic chronometers}
Cosmic chronometers (CC) are part of a method that allows the Hubble parameter to be determined independently of cosmology. This approach is based on the assumption that the expansion of the universe causes the age of  passively evolving galaxies to change with redshift \cite{simon05,stern10,moresco12,moresco15,zhang14,CC2,Borghi_2022}. In this case we employ the same data set used in articles \cite{Plaza:2025gcv} 
\subsection{Supernovae type Ia}
Type Ia supernovae (SN Ia) are very useful tools for determining cosmic distances due to their uniform luminosities, characteristic light curves, and the uniform distribution they have in space. Furthermore, since they constitute extremely energetic events, they can be observed up to  $z \sim$ 2.3 .

The modulus distance can be estimated both from a theoretical expression involving the luminous distance $d_L$,
\begin{equation}
\mu=25+5 \log_{10}(d_L(z)),
\label{distmod}
\end{equation}
\begin{equation}
d_L(z)=(1+z)\int_0^z\frac{c}{H(z')}dz',
\label{distlum}
\end{equation}
and also thanks to SN Ia data.
\begin{equation}
\mu=m_b-M_{b}
\label{mu_obs}
\end{equation}
where $m_b$ and $M_b$ represent the apparent and absolute magnitudes respectively. We consider two datasets. The first consists of 1,590 data points with redshift 
$z
>
0.01$ from the Pantheon Plus compilation  \cite{2022ApJ...938..110B,2022ApJ...938..113S}. The second dataset (Pantheon Plus+SH0ES ) extends the first by incorporating distance modulus measurements used by the SH0ES collaboration \cite{riess2022ApJ...934L...7R}, calibrated using Cepheid variables. These additional data points correspond to redshifts $z<0.01$. Although these latest data have been questioned by several authors \citep{2024arXiv240806153F,peri2024PhRvD.110l3518P}, they have not yet cast off and it is necessary to take them into account if you want to study whether an alternative model alleviates the $H_0$ tension or not.
Furthermore, their incorporation into the statistical analysis allows us to break the degeneracy between the cosmological parameters $M_{b}$ and $H_0$ in SN Ia without having to impose another marginalization on the $M_{b}$ like in \cite{Conley,Camarena2020,Camarena2023,2024MNRAS.527.7626R,2023PhRvD.107f3523C,PhysRevD.109.123514}.
\subsection{Baryon acoustic oscillations}
\label{BAO}
Other valuable tools for estimating cosmological distances are baryon acoustic oscillations (BAO). These arose before the decoupling of matter and radiation, as density perturbations in baryonic matter propagated as oscillating pressure waves through the photon-baryon fluid. This behavior was driven by gravitational potentials that caused matter to cluster, and by interactions with radiation that scattered the baryons. In this case, we use the lastest data from the DESI DR2 release  \citep{DR22025arXiv250314738D}, since  this updated version offers improved measurement accuracy, owing to a significantly larger sample of galaxies and quasars compared to the first release (DR1) \cite{2024arXiv240403002D}. Notably, the enhanced signal-to-noise ratio in the new quasar dataset now allows for separate measurements of the transverse comoving distance $D_M(z)=\int_0^z \frac{c}{H(z')}dz'$
 and the Hubble distance $D_H(z)=\frac{c}{H(z)}$, representing a major advancement—since DR1 only provided the volume-averaged distance $D_V(z) = \left[  D_M^2(z) \frac{c z}{H(z)}\right]^{1/3}$.

\subsection{Quasar x-ray and UV fluxes}

Quasars are considered promising cosmological probes because they are not only among the most luminous objects in the universe, but are also observable at very high redshifts.
Due to the non-linear relationship between their ultraviolet and X-ray emissions, it is possible to convert quasars into standardizable candles by obtaining a distance modulus-redshift relationship \cite{RL2015,RL2019,LR2020}. These emissions are related by the following empirical expression,
\begin{equation}
    \log L_{\rm X} = \gamma \log L_{\rm UV} + \beta ,
    \label{luminosities}
\end{equation}
being $L_{\rm X}$ and $L_{\rm UV}$ the rest-frame monochromatic luminosities at 2 keV and 2500 \r{A} respectively and $\log$ represents the logarithm to
the base 10. Parameters $\gamma$ and $\beta$ are derived from observational data and are expected to be independent of redshift to ensure the robustness of the method \cite{RL2015,RL2019,LR2020}. Furthermore, due to a significant correlation that exist between the parameters governing the quasar luminosity relation and cosmological distances, the luminosity distances inferred from quasar fluxes must first be cross-calibrated, for instance, using Type Ia supernovae as a reference standard.

However, given that it is a relatively recent procedure, it remains subject to scrutiny and ongoing debate. While some authors \cite{KR2021,KR2021_2,PhysRevD.105.103510,10.1093/mnras/stac2325} argue that some of the subsamples in this data collection are not standardizable as they are not model-independent (cosmological) and/or evolve with redshift,  Dainotti et al. \cite{2022ApJ...931..106D} showed that the X-ray/UV relationship in quasars is not influenced by selection biases or by evolution with redshift, but it has an intrinsic dependence on the physics of quasars. Furthermore, in \cite{2023A&A...676A.143S} the fluxes relation is verified as a standard candle using spectroscopic UV and X-ray data instead of photometric data; being much more precise than those commonly used, especially in the case of UV rays. Meanwhile, several authors \cite{2022ApJ...931..106D,2022ApJ...940..174W,2024ApJ...962..103W,2023ApJS..264...46L} have established a higher reliability of the $L_X-L_{UV}$ relations as a cosmological tool by adding a redshift correction
to the quasar luminosity. This fact allows to obtain a three-dimensional and evolutionary version in redshift of the $L_X-L_{UV}$ relation. In \cite{2022ApJ...931..106D}, a modification term $(1+z)^k$ is included and the $k$ coefficient  is determined by the Efron-Petrosian (EP) method \cite{1992ApJ...399..345E}. While in other article \cite{2022ApJ...940..174W}, the three-dimensional and redshift-evolutionary
fluxes relation is obtained with COPULA, a powerful statistical tool developed by \cite{Nelsen2006}. For this present work,  we have decided to take into account the aforementioned modifications, adopting the coefficient calibration carried out by Zhang H. et al. \cite{2024MNRAS.530.4493Z}, based on Type Ia supernova data from the Pantheon+ compilation. In their study, a Gaussian process is employed to estimate the parameters for each approach: the standard one and the two that include a redshift dependence.

The following expression provides a general representation of the relationship between the fluxes of the AGN
\begin{eqnarray}\label{eq:evo_F}
	\log \left(F_{X}\right)&=&2(\gamma-1) \log \left(d_{L}\right)+\beta+(\gamma-1) \log (4 \pi) \nonumber \\
	&& +\gamma \log \left(F_{U V}\right)+\alpha \ln(\bar{\alpha}+z).
\end{eqnarray}
where the relationships with the respective luminosities are given by $F_X=\frac{L_X}{4\pi d_L^2}$ and $F_{UV}=\frac{L_{UV}}{4\pi d_L^2}$, being $d_L$ the luminosity distance.
When the coefficient 
$\alpha$ is zero, the traditional formulation is recovered, whereas nonzero values of $\alpha$ account for redshift evolution. A value of $\bar{\alpha}=1$ corresponds to the modification introduced by \cite{2022ApJ...931..106D}, which we refer to as AGN I, while 
$\bar{\alpha}=5$ belongs to the correction proposed by \cite{2022ApJ...940..174W} (AGN II). The above equation can be rewritten in terms of the parameters $\alpha$, $\beta'$ and $\gamma$ which are the ones estimated in \cite{2024MNRAS.530.4493Z} together with the dispersion $\delta$ found in the likelihood.
\begin{eqnarray}\label{eq:evo_F_reduce}
	\log \left(F_{X}\right)  =2(\gamma-1) \frac{m}{5}+\beta'+\gamma \log \left(F_{U V}\right)+\alpha \ln(\bar{\alpha}+z).
\end{eqnarray}
being $m$ the apparent magnitude,
\begin{eqnarray}\label{eq:mth}
	m=5 \log D_{L} + 25 + M - 5\log \frac{c}{H_0},
\end{eqnarray}
and
\begin{eqnarray}\label{eq:dL}
	D_L=\frac{H_0}{c} d_L
\end{eqnarray}
The maximum likelihood of the following function will be used to estimate the parameters of the cosmological models;
\begin{small}
\begin{eqnarray}\label{eq:likelihoodAGN}
 \tilde{\mathcal{L}} (\bm{p}) &=& \prod_{i=1}^{N} \frac{1}{\sqrt{2 \pi\left(\tilde{\sigma}_{i}^{2}+\delta^{2}\right)}} \times \\ \nonumber 
&&\exp \left\{-\frac{\left[\log \left(F_{X}\right)_{\mathrm{obs},i}-\log \left(F_{X}\right)_\mathrm{th}\right]^{2}}{2\left(\tilde{\sigma}_{i}^{2}+\delta^{2}\right)}\right\}.
\end{eqnarray}
\end{small}
with $N$ the number of data and $\delta$ the intrinsic dispersion \footnote{This parameter is estimated jointly with $\alpha$, $\beta'$ and $\gamma$.}. Later, $\tilde{\sigma}_{i}$ is expressed as,
\begin{small}
\begin{eqnarray}
	\tilde{\sigma}_{i}^{2}&=&\sigma_{{\log \left(F_{X}\right)}_{\mathrm{obs},i}}^{2}+\left(\frac{\partial \log \left(F_{X}\right)_\mathrm{t h}}{\partial \gamma}\right)_{i}^{2} \sigma_{\gamma}^{2} \nonumber\\ &+&\left(\frac{\partial \log \left(F_{X}\right)_\mathrm{t h}}{\partial \beta'}\right)_{i}^{2} \sigma_{\beta'}^{2}
	+\left(\frac{\partial \log \left(F_{X}\right)_\mathrm{t h}}{\partial \alpha}\right)_{i}^{2} \sigma_{\alpha}^{2} \nonumber\\ &+&\left(\frac{\partial \log \left(F_{X}\right)_\mathrm{t h}}{\partial \log \left(F_{U V}\right)_\mathrm{obs}}\right)_{i}^{2} \sigma_{\log \left(F_{U V}\right)_\mathrm{obs}}^{2}\nonumber\\
	&+&2 \sum_{j=1}^3 \sum_{k=j+1}^3\left(\frac{\partial \log \left(F_{X}\right)_\mathrm{t h}}{\partial {\theta}_{j}} \frac{\partial \log \left(F_{X}\right)_\mathrm{t h}}{\partial {\theta}_{k}}\right)_{i} C_{j k},
\end{eqnarray}
\end{small}
where  $\bm{\theta}\equiv\left\{\alpha,~\beta',~\gamma \right\}$, $C_{j k}$ represents the covariance matrix, and $\sigma_\alpha$,
$\sigma_{\beta'}$, and $\sigma_\gamma$ are the uncertainties of coefficients $\alpha$, $\beta'$, and $\gamma$ obtained from the gaussian process \cite{2024MNRAS.530.4493Z}. Although the absolute magnitude $M_b$ and $H_0$
  cannot be simultaneously constrained using AGN data alone using this kind of parameter calibration \cite{2024MNRAS.530.4493Z},  this limitation is not critical in our case, as additional datasets, such as cosmic chronometers, SH0ES, etc. are incorporated to break this degeneracy. 

We consider the compilation of 2421 X-ray and UV flux measurements of quasars quasi-stellar objects (QSOs/AGN) published by  Lusso et al. \cite{LR2020} that covers a redshift range of  $0.009 \leq z \leq 7.5413$. The later has already been used to constrain other cosmological models \cite{M2022PhRvD.105j3526L,2024arXiv241204830Z,2024EPJC...84..444L} and many others.


\subsection{Cosmic Microwave Background}

The CMB provides a rich source of physical information about the early Universe. For a given cosmological model and a set of initial conditions, the angular power spectrum of the CMB anisotropies can be computed numerically to high precision using linear perturbation theory. The final \textit{Planck} 2018 data release \cite{Planckcosmo2018,Planck2018overview} provides high-precision measurements of the CMB angular power spectra, namely the temperature anisotropy auto-spectrum (TT), the E-mode polarization auto-spectrum (EE), the temperature–polarization cross-spectrum (TE), and the CMB lensing potential power spectrum. The combination of precise experimental measurements and accurate theoretical predictions allows one to set tight constraints on cosmological parameters. For the standard $\Lambda$CDM model, these include: the dark matter and baryon density parameters $\Omega_{cdm}h^2$ and $\Omega_{b} h^2$ (with $h \equiv H_0/100$ [km/s/Mpc]), the angular acoustic scale at recombination $100\,\theta_s$, the reionization optical depth $\tau_{\rm{reio}}$, the amplitude of the primordial scalar power spectrum $\log(10^{10}A_s)$, and the scalar spectral index $n_s$.

Assuming statistical isotropy, the CMB multipole coefficients $a_{\ell m}^X$ (for $X = T$ or $E$) satisfy 
\begin{equation}
	\langle a_{\ell m}^X a_{\ell' m'}^{* Y} \rangle = C_\ell^{XY} \,\delta_{\ell\ell'}\delta_{mm'}, 
\end{equation}
so that the TT, TE, and EE spectra ($C_\ell^{TT}$, $C_\ell^{TE}$, $C_\ell^{EE}$), together with the lensing spectrum, succinctly characterize the two-point statistics of the CMB anisotropies in harmonic space. In the \textit{Planck} 2018 release \cite{Planck2018like}, these measured spectra (and their covariances) are treated as the observational inputs to the likelihood, condensing the information from CMB sky maps into a form that can be efficiently compared with theoretical models.

In our analysis, we incorporate the CMB temperature (TT) and polarization (TE, EE) spectra from the \textit{Planck} collaboration by employing the \texttt{simall} and \texttt{Commander} likelihoods \cite{Planck2018like} for $\ell < 30$, and the \texttt{CamSpec} likelihood for $\ell \geq 30$. The \texttt{CamSpec} likelihood \cite{camspec}, based on the latest \texttt{NPIPE PR4} data release from \textit{Planck}, replaces the original high-$\ell$ \texttt{Plik} likelihood \cite{Planck2018like}. It is important to mention that these same likelihoods are also used in the analysis of the DESI DR2 collaboration \cite{DR22025arXiv250314738D}. Additionally, we adopt the \textit{Planck} \texttt{NPIPE PR4} lensing likelihood together with the \textit{Planck} \texttt{PR4} ISW-lensing likelihoods \cite{carron2022}.

\section{Results and Discussions}
\label{secIV}

We present the results of our statistical analysis for the UG model described in Section \ref{secII} applying the observational data reported in Section \ref{secIII}, and the results for the standard  $\Lambda$CDM model for comparison. 

The posterior probability distributions were obtained using the publicly available Bayesian inference code \texttt{COBAYA} \cite{cobaya1,cobaya2}. This code enables us to probe the UG cosmological model by confronting it with the experimental likelihoods and exploring the parameter space through a MCMC sampler algorithm \cite{mcmcsampler1,mcmcsampler2}. This method implements a fast-dragging subroutine that enables efficient oversampling to accelerate the sampling of parameters corresponding to experimental nuisance quantities. The predictions for the UG cosmological model were computed with a customized version of the \texttt{CLASS} Boltzmann solver \cite{class}. The modifications to \texttt{CLASS} included the new evolution equations for the baryon and cold dark matter energy densities, Eqs.(\ref{rhomsolb}) and (\ref{rhomsolcdm}), together with the evolution equation for the diffusion term $Q$, Eq.(\ref{Qfinal}).

The free parameters to be estimated are: $\beta$, which quantifies\footnote{Recall that if $\beta=1$, our model is equivalent to the $\Lambda$CDM model. } the effect of the violation of the continuity equation for $\rho_m$ encoded in $Q$ (equivalently $\beta$ also reflects the strength of the interaction between $\rho_H$ and $\rho_m$); $H_0$, the Hubble parameter; $\Omega_{cdm}h^2$ and $\Omega_{b} h^2$ represent cold dark matter and baryonic matter densities respectively, where $h=H_0/100$[km/s/Mpc]. The absolute magnitude $M_b$ of the SN Ia has been also treated as a free parameter. When the CMB  dataset is included, we analyze the same parameters $\beta$ and $M_b$ together with the six standard  $\Lambda$CDM parameters: $\Omega_{cdm}h^2$ and $\Omega_{b} h^2$, the angular acoustic scale at recombination $100 \theta_s$, the reionization optical depth $\tau_{\rm{reio}}$, the primordial scalar power
spectrum amplitude $\log(10^{10}A_s)$, and the scalar spectral index $n_s$. Table \ref{tabla0} summarizes the flat priors assigned to the free parameters considered in our study. These ranges are sufficiently broad to include all statistically confident values of the model parameters. We set the Gelman-Rubin statistic criterion for chain convergence at $R-1 < 0.05$.

\begin{figure*}[!ht]
    \centering
    \includegraphics[width=0.9\linewidth]{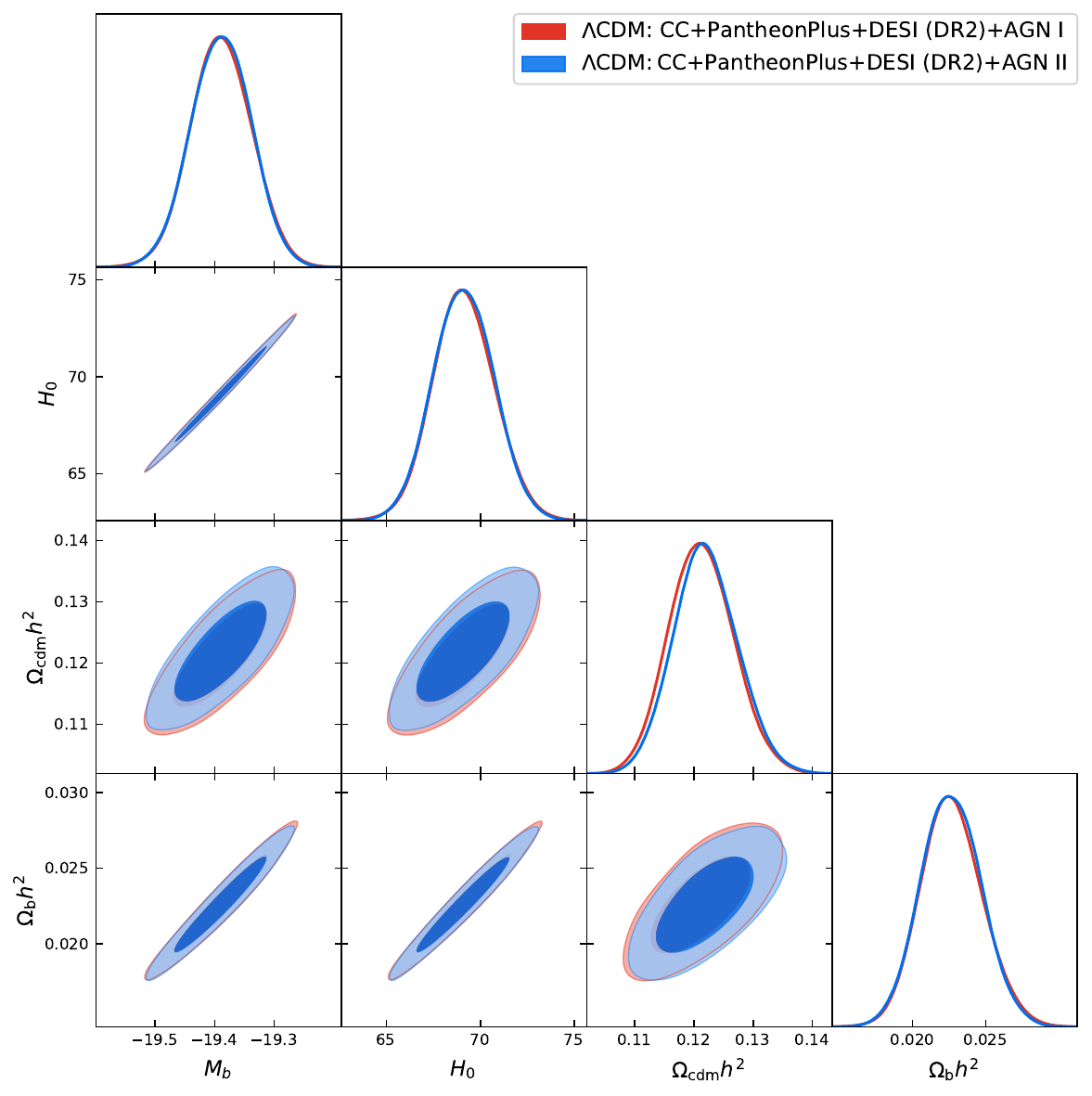}
    \caption{Comparison of parameter estimation between the two QSO/AGN models, labeled AGNI and AGNII, within the $\Lambda$CDM cosmological framework. The posterior probability distributions for each parameter are shown, along with the 68\% and 95\% confidence level contours. No significant differences are observed between the two AGN models.}
    \label{fig1:1y2lcdm}
\end{figure*}

\begin{figure*}[!ht]
    \centering
    \includegraphics[width=0.9\linewidth]{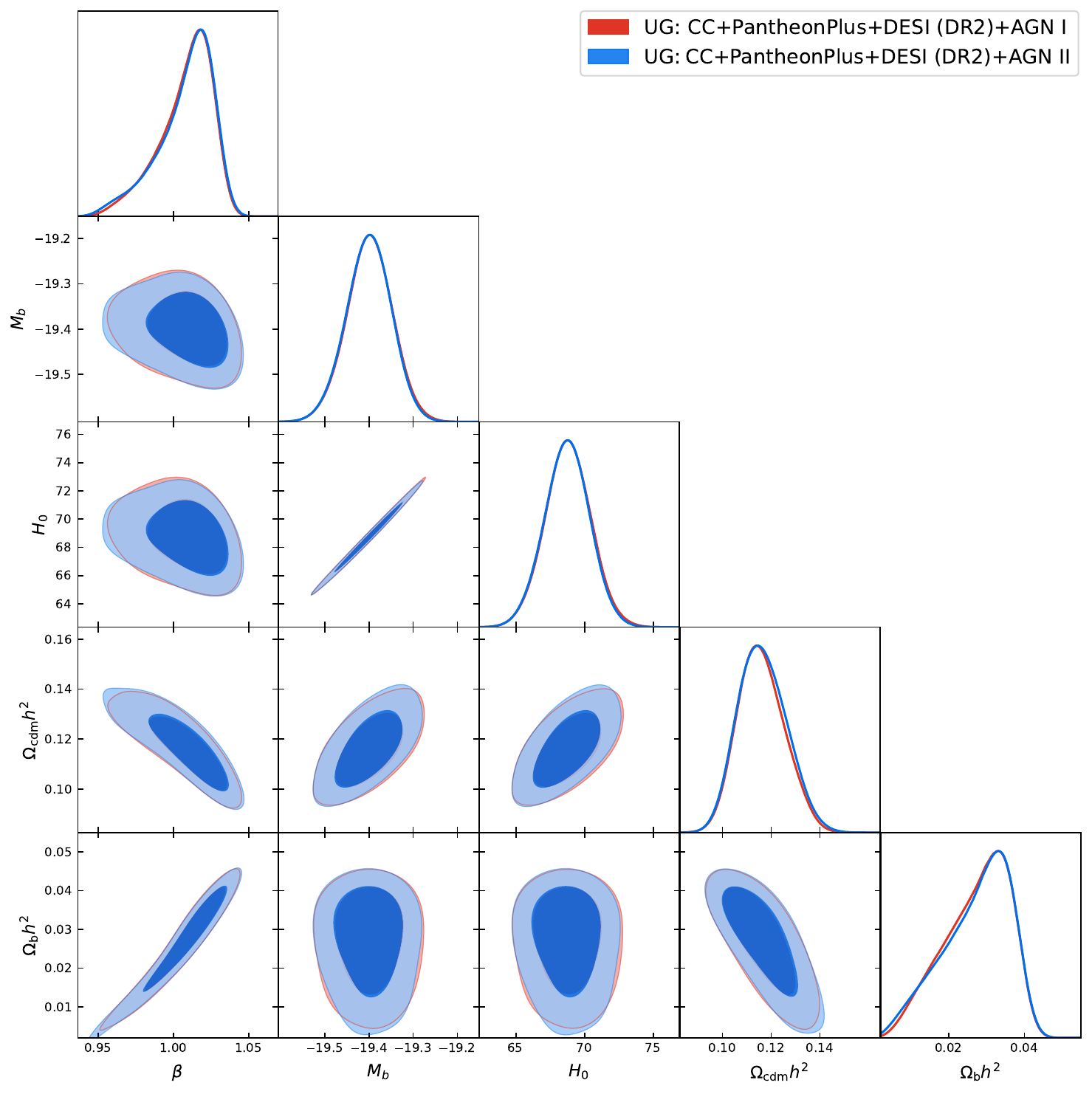}
    \caption{Comparison of parameter estimation between the two QSO/AGN models, labeled AGNI and AGNII, within the cosmological model based on UG. The posterior probability distributions for each parameter are shown, along with the 68\% and 95\% confidence level contours. As in the previous case, no significant differences are observed between the two AGN models.}
    \label{fig2:1y2velten}
\end{figure*}

From Figures \ref{fig1:1y2lcdm} and \ref{fig2:1y2velten}, as well as Tables \ref{tabla6} and \ref{tabla7}, it can be observed that, for both the $\Lambda$CDM and UG models, the parameter estimates exhibit minimal variation when either the AGNI or AGNII approach is applied to the quasar data. For this reason, only one AGN approach is employed in the subsequent analyses, namely the one proposed by Dainotti et~al.\ \cite{2022ApJ...931..106D}. Moreover, for the late Universe data sets considered, the results of the two cosmological models are consistent within $1\sigma$.

\begin{figure*}[!ht]
    \centering
    \includegraphics[width=0.9\linewidth]{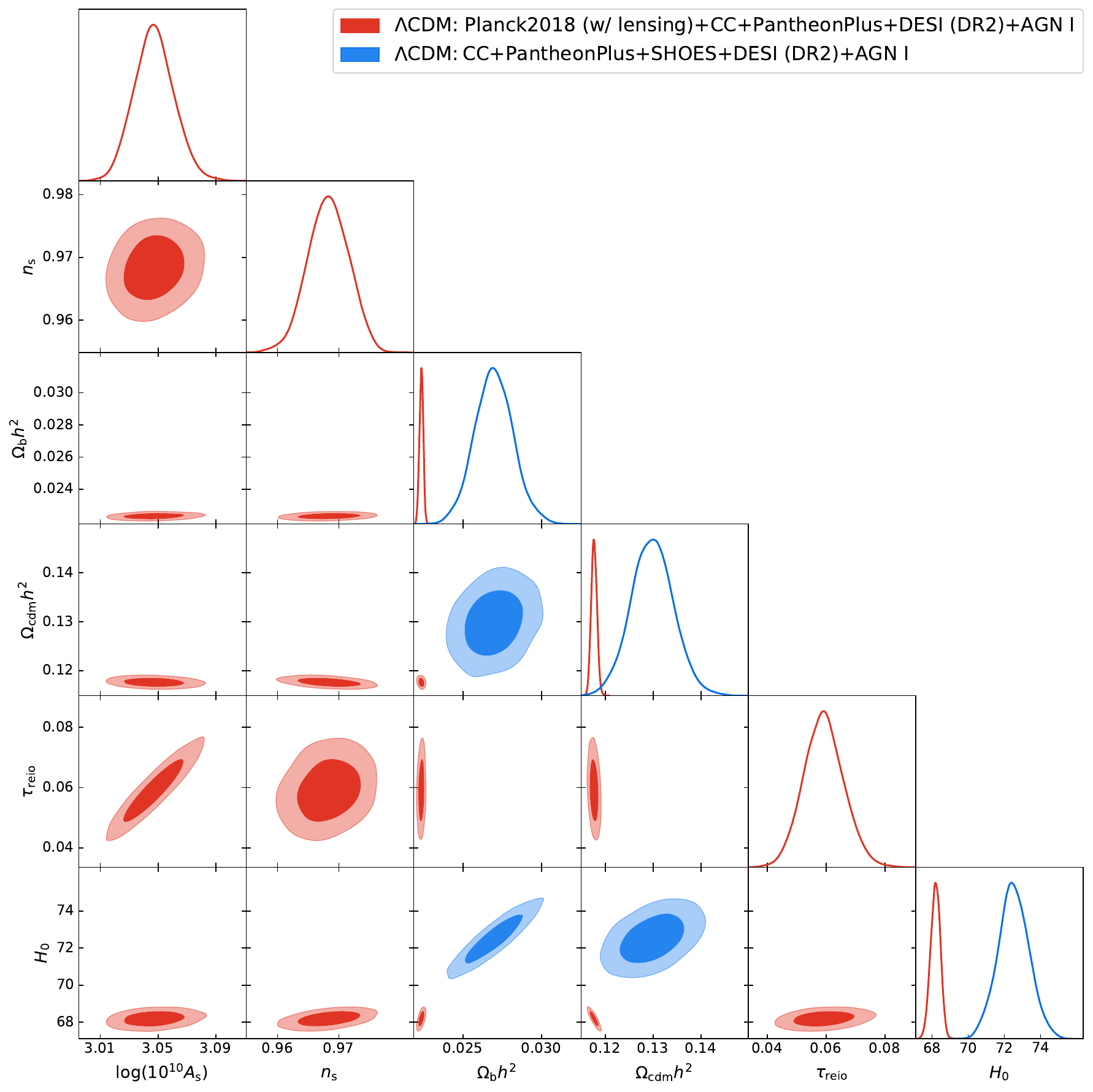}
    \caption{Posterior probability distributions and the 68\% and 95\% confidence regions for the six standard cosmological parameters within the $\Lambda$CDM model. We employ both early- (red plots) and late-time (blue plots) Universe datasets. The $H_0$ tension is evident.}
    \label{fig3:3y4lcdm}
\end{figure*}

\begin{figure*}[!ht]
    \centering
    \includegraphics[width=0.9\linewidth]{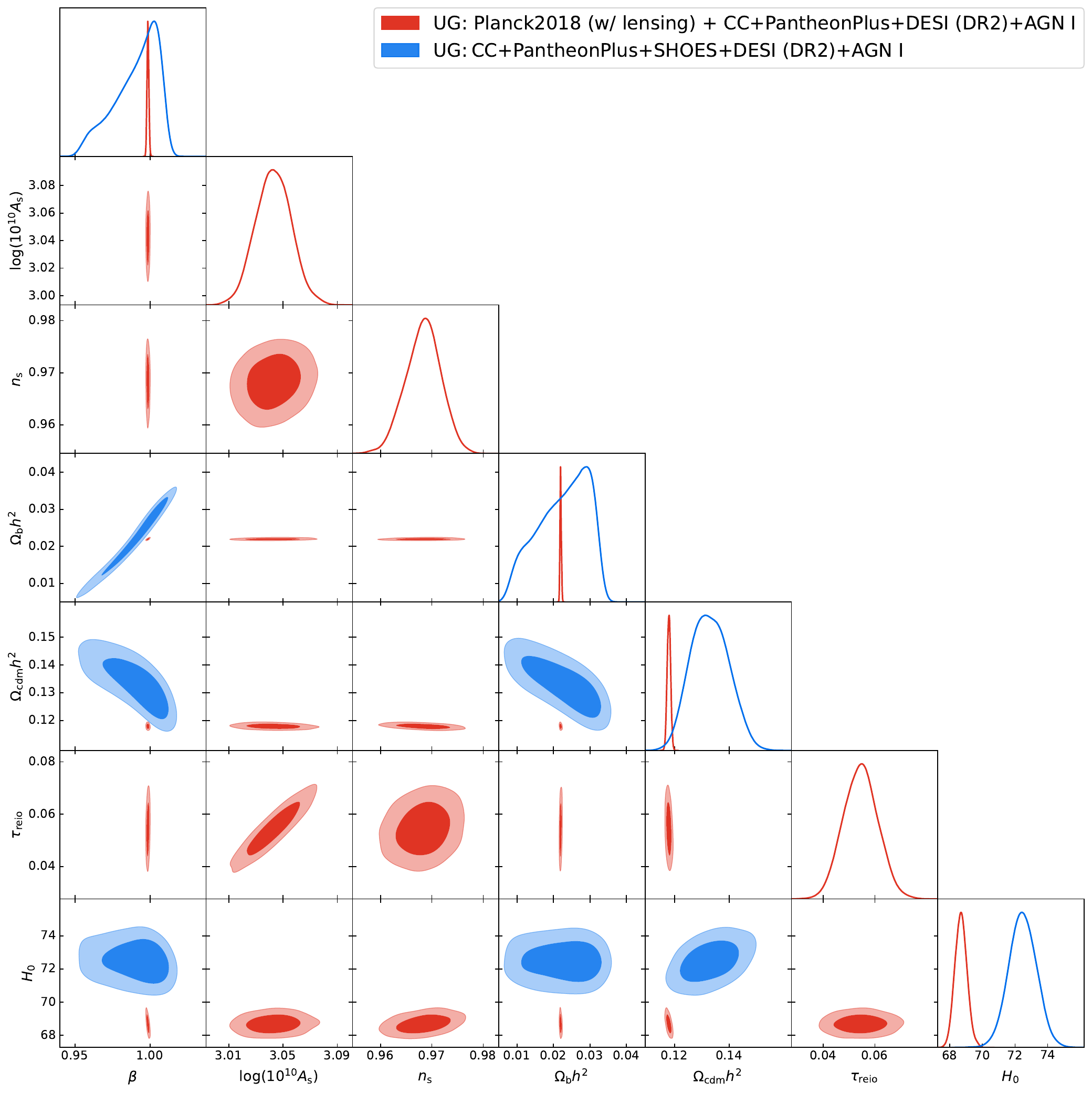}
    \caption{Posterior probability distributions and the 68\% and 95\% confidence regions for the $\beta$ parameter and the standard 6 cosmological parameters within the UG cosmological model. We employ both early- (red plots) and late-time (blue plots) Universe datasets. The $H_0$ tension is not alleviated in this model.}
    \label{fig4:3y4velten}
\end{figure*}

However, when CMB and/or SH0ES data are added to the late Universe datasets, the resulting parameter estimates differ significantly, for a given cosmological model, between the two theoretical frameworks. Tables \ref{tabla6}, \ref{tabla7}, \ref{tabla8}, and \ref{tabla9}, together with Figures \ref{fig1:1y2lcdm}, \ref{fig2:1y2velten}, \ref{fig3:3y4lcdm}, and \ref{fig4:3y4velten}, indicate that, for both models, the constraining power of the early-Universe data is substantially greater. The intervals for the $\beta$ parameter obtained from the various statistical analyses are mutually consistent and compatible with the standard model at the 1$\sigma$ level. Despite that, when the Planck2018 (w/ lensing) data are included, this consistency is achieved only at the 3$\sigma$ level. On the other hand, the absolute magnitude estimates from analyses incorporating the Planck2018 (w/ lensing) data and those from the SH0ES collaboration are mutually consistent only at the 3$\sigma$ level, whereas the estimate presented in the left column Table \ref{tabla7} is compatible with both at the 1$\sigma$ level. This discrepancy is also reflected in the $H_0$ estimates, where the tension exceeds the 3$\sigma$ level. From this comparison, it is clear that, like the $\Lambda$CDM model, this unimodular gravity model is unable to alleviate the Hubble tension. The remaining parameters $\Omega_bh^2$ and $\Omega_{cdm}h^2$ are consistent at 1 or 2$\sigma$, depending on the case. Finally, when comparing the estimated parameter intervals obtained from each dataset with those derived from the $\Lambda$CDM model, no significant differences are observed, with most of them being consistent at the 1$\sigma$ and/or 2$\sigma$ level.

\subsection{Statistical Criteria for Model Selection}

To determine which of the models presented in this work is statistically preferred for the selected dataset, we employ several statistical criteria. The minimum chi-square value ($\chi^2_{min}$) is directly related to the maximum likelihood estimate, while the reduced chi-square, $\chi^2_{\nu}$, is defined as $\chi^2_{min}/\nu$, where $\nu=N-k$ denotes the number of degrees of freedom, with $N$ being the total number of data points and $k$ the number of free parameters in the model. We also consider the Akaike Information Criterion (AIC) \cite{1974ITAC...19..716A}  and the Bayesian Information Criterion (BIC) \cite{10.1214/aos/1176344136}, \cite{doi:10.1177/0049124104268644}, given by:
\begin{equation}
   { \rm AIC}=-2 \ln \mathcal{L}_{max}+2k
\end{equation}
\begin{equation}
   { \rm BIC}=-2 \ln \mathcal{L}_{max}+k\ln N 
\end{equation}
Statistically, the model that best fits the data is the one with the lowest values of $\chi^2_{min}$, AIC, and BIC. However, in the case of the reduced chi-square, the most favored model is the one with a $\chi^2_\nu$ value closest to 1. A value of $\chi^2_{\nu}>>1$ suggests underfitting—indicating that the model fails to capture the data adequately—whereas $\chi^2_{\nu}<<1$ points to overfitting, either due to the model fitting noise or due to overestimated error variances \cite{Bevington1969}. Among these criteria, $\chi^2_{min}$ is the only one that does not account for the number of free parameters in the model. This is a crucial distinction, as in nested models, the likelihood generally increases with the number of parameters, irrespective of their actual relevance. In contrast, AIC and BIC incorporate penalties for model complexity, albeit in different ways—with BIC applying the stricter penalty. This difference can lead to disagreements among the criteria, particularly when the underlying assumptions of each are violated (\cite{doi:10.1177/0049124104268644}, \cite{2004MNRAS.351L..49L} among others). Comparative model assessment using AIC or BIC is performed by evaluating the difference in values between two models ($\Delta$X = $\Delta$ AIC or $\Delta$X = $\Delta$BIC), interpreted as follows:
\begin{itemize}
    \item{ $0\leq\Delta X\leq 2$ or $-2 \leq \Delta X < 0 $:
 weak evidence; no clear preference}
   
    \item{$2 < \Delta X \leq 6$ or $-6 \leq \Delta X < -2$: positive evidence}
    \item{$6 < \Delta X \leq 10$ or $-10 \leq \Delta X < -6$: strong evidence}
    \item{$\Delta X > 10$ or $\Delta X < -10$: very strong evidence}
\end{itemize}
Based on the results presented in Table \ref{tabla10} and the preceding discussion, it is evident that the data combination CC+PantheonPlus+SH0ES +DESI (DR2)+AGN I yields higher likelihoods for the models under analysis compared to the combination Planck2018 +CC+PantheonPlus +DESI (DR2)+AGN I. Notably, this case exhibits potential signs of either overfitting or overestimated uncertainties in the AGN data, as evidenced by the low reduced chi-square value of $\chi^2_{\nu} \approx 0.7$, which arises from the inclusion of these recent and relatively unexplored observations, where error overestimations remain plausible. In previous studies such as \citep{Plaza:2025gcv}, which used the same dataset except for the AGN, the standard model did not yield such a low $\chi^2_{\nu}$. 
Regarding the study of Bayesian information criteria, positive (AIC) and strong (BIC) evidence is shown in favor of $\Lambda$CDM when the SH0ES dataset is included. Meanwhile, applying the other data combination, there is a positive preference for the UG model using the AIC criteria, and a positive preference for the $\Lambda$CDM model when the BIC is used. 

\begin{table}[!ht]
	\centering
	\begin{tabular} { c c }
		\hline
		\hline
	\textbf{Parameter} & \textbf{Prior} \\
		\hline
		\hline 
        $\beta$   & $[0.4,2.5]$ \\
        $M_b$ & $[-20,-18]$ \\
        $H_0$ [km/s/Mpc] & $[20,100]$ \\
        $\Omega_{\rm b} h^2$ & $[0.005,0.1]$ \\
        $\Omega_{\rm cdm} h^2$ & $[0.001,0.99] $ \\
        $\tau_\mathrm{reio}$ & $ [0.01,0.8] $ \\
        $100\theta_s $ & $[0.5,10] $ \\
        $\log(10^{10} A_s)$ & $[1.61,3.91] $  \\
        $n_s$ & $[0.8,1.2]$ \\
        \hline
		\hline
	\end{tabular}
	\caption{ Flat priors on the model parameters used
		during the statistical analysis. }
	\label{tabla0}
\end{table}

\begin{table*}[!ht]
	\centering
	\begin{tabular} {c c c  c c }
		\hline
		\hline
		& $ $ & \textbf{ Model: $\Lambda$CDM }  & $ $ & $ $ \\
		\hline
		$ $ &  \emph{ CC+PantheonPlus }  & $ $ &  \emph{CC+PantheonPlus} & $ $ \\
		$ $ & \textit{+DESI (DR2)+AGN I} & $ $ & \textit{+DESI (DR2)+AGN 
			II} & $ $ \\
		\hline
		Parameter & Mean value and 68 \% CL & Best fit & Mean value and 68 \% CL & Best fit \\
		\hline
	{\boldmath$M_b            $} & $-19.389\pm 0.051          $& $-19.394 $  & $-19.389\pm 0.050          $ & $-19.398$ \\
	
	{\boldmath$H_0$ [km/s/Mpc] } & $69.1\pm 1.6               $ & 68.94 & $69.1\pm 1.6               $ & 68.8\\
	
	{\boldmath$\Omega_\mathrm{cdm} h^2$} & $0.1213\pm 0.0054          $ &  0.1203  & $0.1220\pm 0.0054          $& 0.1203 \\
	
	{\boldmath$\Omega_\mathrm{b} h^2$} & $0.0227^{+0.0019}_{-0.0022}$ &  0.0225 & $0.0226\pm 0.0021          $ & 0.0222\\
		\hline
		\hline
	\end{tabular}
	\caption{$\Lambda$CDM results from statistical analysis using different datasets combinations (Cosmic Chronometers (CC), SneIa from
		PantheonPlus, BAO dataset from DESI (DR2) collaboration, and Quasars employing AGN I and AGN II). For each parameter, we present the mean value and the 68\% confidence levels or the upper/lower limits obtained.}
	\label{tabla6}
\end{table*}

\begin{table*}[!ht]
	\centering
	\begin{tabular} {c c c  c c }
		\hline
		\hline
		& $ $ & \textbf{ Model: Unimodular Gravity}  & $ $ & $ $ \\
		\hline
		$ $ &  \emph{ CC+PantheonPlus }  & $ $ &  \emph{CC+PantheonPlus} & $ $ \\
		$ $ & \textit{+DESI (DR2)+AGN I} & $ $ & \textit{+DESI (DR2)+AGN 
			II} & $ $ \\
		\hline
		Parameter & Mean value and 68 \% CL & Best fit & Mean value and 68 \% CL & Best fit \\
		\hline
			{\boldmath$\beta          $} & $1.009^{+0.022}_{-0.0092}  $ & 1.027 & $1.009^{+0.022}_{-0.0087}  $& 1.029\\
		
		{\boldmath$M_b            $} & $-19.399\pm 0.051          $& $-19.401$ & $-19.401\pm 0.051          $ & $-19.402$\\
		
		{\boldmath$H_0            $ [km/s/Mpc]} & $68.8\pm 1.7               $& 68.6& $68.7\pm 1.6               $& 68.6\\
		
		{\boldmath$\Omega_\mathrm{cdm} h^2$} & $0.1159^{+0.0085}_{-0.011} $& 0.1070 & $0.1164^{+0.0090}_{-0.011} $& 0.1066\\
		
		{\boldmath$\Omega_\mathrm{b} h^2$} & $0.0276^{+0.012}_{-0.0043} $& 0.0381& $0.0276^{+0.012}_{-0.0042} $& 0.0391\\
		\hline
		\hline
	\end{tabular}
	\caption{UG results from statistical analysis using different datasets combinations (Cosmic Chronometers (CC), SneIa from
		PantheonPlus, BAO dataset from DESI (DR2) collaboration, and Quasars employing AGN I and AGN II). For each parameter, we present the mean value and the 68\% confidence levels or the upper/lower limits obtained.}
	\label{tabla7}
\end{table*}

\begin{table*}[!ht]
	\centering
	\begin{tabular} {c c c  c c }
		\hline
		\hline
			& $ $ & \textbf{ Model: $\Lambda$CDM }  & $ $ & $ $ \\
		\hline
		$ $ &  \emph{ Planck2018 +CC+PantheonPlus }  & $ $ &  \emph{CC+PantheonPlus+SH0ES} & $ $ \\
		$ $ & \textit{+DESI (DR2)+AGN I} & $ $ & \textit{+DESI (DR2)+AGN I} & $ $ \\
		\hline
		Parameter & Mean value and 68 \% CL & Best fit & Mean value and 68 \% CL & Best fit \\
		\hline
	{\boldmath$M_b            $} & $-19.4173\pm 0.0083        $ & $-19.4152$ & $-19.284\pm 0.026          $ & $-19.2834$ \\

{\boldmath$\log(10^{10} A_\mathrm{s})$} & $3.047\pm 0.014            $ & 3.047 & -- & -- \\

{\boldmath$n_\mathrm{s}   $} & $0.9683\pm 0.0034          $& 0.9687 & -- & --\\

{\boldmath$100\theta_\mathrm{s}$} & $1.04193\pm 0.00023        $ & 1.04195& -- & -- \\

{\boldmath$\Omega_\mathrm{b} h^2$} & $0.02233\pm 0.00012        $ & 0.02232& $0.0270\pm 0.0012          $ & 0.0270 \\

{\boldmath$\Omega_\mathrm{cdm} h^2$} & $0.11765\pm 0.00059        $& 0.11748 &  $0.1298\pm 0.0045          $ & 0.1292\\

{\boldmath$\tau_\mathrm{reio}$} & $0.0592\pm 0.0069          $ &0.0615 & -- & -- \\

$H_0                       $ [km/s/Mpc] & $68.19\pm 0.27             $& 68.24& $72.51\pm 0.87             $ & 72.54 \\
		\hline
		\hline
	\end{tabular}
	\caption{$\Lambda$CDM results from statistical analysis using different datasets combinations (CMB from \textit{Planck 2018} collaboration, Cosmic Chronometers (CC), SneIa from
		PantheonPlus+SH0ES, BAO dataset from DESI (DR2) collaboration, and Quasars employing AGN I). For each parameter, we present the mean value and the 68\% confidence levels or the upper/lower limits obtained.}
	\label{tabla8}
\end{table*}

\begin{table*}[!ht]
	\centering
	\begin{tabular} {c c c  c c }
		\hline
		\hline
		& $ $ & \textbf{ Model: Unimodular Gravity }  & $ $ & $ $ \\
		\hline
		$ $ &  \emph{ Planck2018 +CC+PantheonPlus }  & $ $ &  \emph{CC+PantheonPlus+SH0ES} & $ $ \\
		$ $ & \textit{+DESI (DR2)+AGN I} & $ $ & \textit{+DESI (DR2)+AGN I} & $ $ \\
		\hline
		Parameter & Mean value and 68 \% CL & Best fit & Mean value and 68 \% CL & Best fit \\
		\hline
			{\boldmath$\beta$} & $0.99858\pm 0.00061 $ & 0.99819   & $0.991^{+0.019}_{-0.0076}  $ & 1.00912\\
			
			{\boldmath$M_b            $} & $-19.403\pm 0.011 $ & $-19.399$  & $-19.286\pm 0.025          $& $-19.289$\\
			
			{\boldmath$\log(10^{10} A_\mathrm{s})$} & $3.043\pm 0.013            $ & 3.036 & -- & -- \\
			
			{\boldmath$n_\mathrm{s}   $} & $0.9683^{+0.0036}_{-0.0033}$& 0.9688 & -- & -- \\
			
			{\boldmath$100\theta_\mathrm{s}$} & $1.04195\pm 0.00023        $& 1.04177 & -- & -- \\
			
			{\boldmath$\Omega_\mathrm{b} h^2$} & $0.02197^{+0.00016}_{-0.00021}$&  0.02181 & $0.0227^{+0.0099}_{-0.0038}$ & 0.0323\\
			
			{\boldmath$\Omega_\mathrm{cdm} h^2$} & $0.11790\pm 0.00062        $&  0.1178 & $0.1329^{+0.0067}_{-0.0075}$& 0.1252\\
			
			{\boldmath$\tau_\mathrm{reio}$} & $0.0545\pm 0.0067          $& 0.0493 & -- & -- \\
			
			$H_0                       $ [km/s/Mpc] & $68.69\pm 0.37             $& 68.78  & $72.47\pm 0.85             $& 72.39\\
			\hline
	
		\hline
		\hline
	\end{tabular}
	\caption{UG results from statistical analysis using different datasets combinations (CMB from \textit{Planck 2018} collaboration, Cosmic Chronometers (CC), SneIa from
		PantheonPlus+SH0ES, BAO dataset from DESI (DR2) collaboration, and Quasars employing AGN I). For each parameter, we present the mean value and the 68\% confidence levels or the upper/lower limits obtained.}
	\label{tabla9}
\end{table*}

\begin{table*}[!ht]
	\centering
	\begin{tabular} {c c c  c c  c c c}
\hline
\hline
Data & Model & $\chi^2_{\rm min}$ & $\chi^2_{\nu}$ & AIC & BIC & $\Delta$AIC & $\Delta$BIC \\
\hline
\emph{ Planck2018 +CC+PantheonPlus } & $\Lambda$CDM & 13880.290 & 0.9827 & 13894.290 & 13947.183 & -- & -- \\
\textit{+DESI (DR2)+AGN I} & UG & 13875.086 & 0.9824 & 13891.086 & 13951.535& $-3.204$ & 4.352 \\  
\hline 
\emph{CC+PantheonPlus+SHOES}  & $\Lambda$CDM & 2953.523 & 0.7084 & 2961.523 & 2986.868 & -- & -- \\
\textit{+DESI (DR2)+AGN I}  & UG & 2954.654 & 0.7089 & 2964.654 & 2996.336 & 3.132 & 9.468   \\ 
		\hline
        \hline
	\end{tabular}
	\caption{Results of standard statistical tools commonly used to assess model fitting: the chi-square and the
reduced chi-square statistics $\chi^2_{\rm min}$ and $\chi^2_{\nu}$, the Akaike Information Criterion (AIC), and the Bayesian Information Criterion (BIC) for each model.}
	\label{tabla10}
\end{table*}

\section{Summary and Conclusions}
\label{secV}

In this article, we investigate a cosmological model based on an alternative theory of gravity known as Unimodular Gravity (UG). Specifically, we analyze the model proposed in \cite{velten2024}, which incorporates an interaction between matter and holographic dark energy within the UG framework. We generalize this model to include radiation, which had not been taken into account previously, and to distinguish between baryonic and cold dark matter. Moreover, we employ datasets from both the early and late Universe to evaluate the ability of the UG cosmological model to account for observations across different cosmological epochs, and to assess its potential to alleviate the Hubble tension between the measurements of Riess et al. \citep{3riess2024IAUS..376...15R} and the Planck Collaboration \cite{Planckcosmo2018}.

The analysis includes data from Cosmic Chronometers (CC), Type Ia Supernovae from the Pantheon Plus (PP) and Pantheon Plus + SH0ES (PPS) compilations, Active Galactic Nuclei (AGN), Baryon Acoustic Oscillations from the DESI DR2 project (BAO), and Cosmic Microwave Background (CMB) data from the \textit{Planck} 2018 mission. Constraints on the free parameters were derived for both the UG cosmological model and the standard $\Lambda$CDM model. Additionally, a Bayesian model comparison was performed between them.

Regarding the AGN dataset, two distinct methodologies were employed, as described in Section \ref{secIII}, to evaluate the impact of methodological choices on the results. For this purpose, two statistical analyses were performed using the datasets CC+PP+BAO, supplemented with AGNI in the first case and AGNII in the second. Both analyses were carried out for the standard and UG models. The results indicate that the two AGN methodologies yield practically identical fits in both cases. This is illustrated in Figures \ref{fig1:1y2lcdm} and \ref{fig2:1y2velten}, as well as in Tables \ref{tabla6} and \ref{tabla7}. Consequently, it was determined that the choice of AGN methodology does not significantly affect the results, and the AGNI dataset was arbitrarily selected for the remainder of the analysis.

Subsequently, we performed two additional statistical analyses using different data combinations: CC+PPS+BAO+AGNI and CC+PP+BAO+CMB+AGNI. The first combination emphasizes late universe datasets,  including the SH0ES measurements. The second includes early universe data from the CMB as reported by the \textit{Planck} 2018 collaboration, but excludes SH0ES; hence, placing greater emphasis on early universe constraints. This strategy was designed to probe the potential of the UG model to resolve the Hubble tension. However, we conclude that the UG cosmological model is unable to resolve the Hubble tension, as the estimates obtained from CMB data are inconsistent with the model independent determination by Riess et al. \citep{3riess2024IAUS..376...15R} at more than 3$\sigma$. Similarly, regarding the model-dependent estimates of the present work, the comparison between the CC+PP+AGNI+BAO+CMB and CC+PPS+AGNI+BAO fits reveals a tension in the $H_0$ parameter exceeding $3\sigma$, as shown in Figure \ref{fig4:3y4velten}.

With respect to the parameter $\beta$ of the UG cosmological model, which also characterizes deviations from the standard model ($\beta = 1$ corresponds to $\Lambda$CDM), the best-fit value obtained from the CC+PPS+BAO+AGNI dataset is $\beta = 1.00912$. This indicates a slight departure from $\Lambda$CDM while remaining consistent at the $1\sigma$ level. For the CMB+CC+PP+BAO+AGNI dataset, the best-fit value is also close to unity but inconsistent with the standard value at the $1\sigma$ confidence level (see Table \ref{tabla9}).

For all remaining free parameters common to both the UG and standard cosmological models, the inferred values are very similar and remain mutually consistent within $1\sigma$ for all considered data combinations. This is expected, since a model that requires $\beta$ to be close to 1 to achieve a successful fit naturally reproduces the behavior of the standard model. In particular, as shown in Fig. \ref{fig4:3y4velten} and Table \ref{tabla9}, the parameter $\beta$ cannot depart significantly from unity if the UG model is to remain consistent with the CMB+CC+PP+BAO+AGNI data set. We have verified\footnote{This supplementary analysis is not included here because the results in Fig. \ref{fig4:3y4velten} and Table \ref{tabla9} already imply the same conclusion.} that even small deviations from unity, for example $\beta = 0.97$ or $\beta = 1.05$, while fixing the remaining cosmological parameters at their best-fit values, cause the CMB temperature and polarization spectra ($C_l^{TT}$, $C_l^{TE}$, and $C_l^{EE}$) to strongly diverge from the standard prediction. Consequently, the model fails to reproduce the observations without further testing.


Concerning the Bayesian analysis, for the CMB+CC+PP+BAO+CMB+AGNI dataset, a statistical preference for the UG model is found according to the $\Delta \mathrm{AIC}$, but not according to the $\Delta \mathrm{BIC}$, which instead favors the $\Lambda$CDM. This is reasonable given that the BIC imposes a stronger penalty on models with additional free parameters, in this case the extra parameter $\beta$. However, in both cases the statistical preference is weak. The fit quality is good, with a reduced chi-square value $\chi^2_{\nu} \approx 0.98$.
In contrast, for the dataset CC+PPS+BAO+AGNI, the standard model is statistically favored according to both the $\Delta \mathrm{AIC}$ and $\Delta \mathrm{BIC}$ criteria. While the preference is not strong under the AIC, it is strong according to the BIC. 
In our view, the statistical preference for the $\Lambda$CDM model over the UG cosmological model can be attributed to three key factors: (i) the UG model has one extra free parameter; (ii) the DESI DR2 data tend to favor cosmological models with a time-varying dark energy equation of state, a feature not present in the UG case; and (iii) this work demonstrates that, to successfully reproduce the CMB data, deviations from the standard model must remain minimal.

In conclusion, the cosmological model based on UG explored in this work remains a viable framework for describing both late and early universe observations, including the CMB data. Although the standard model currently has a slight statistical preference, this preference is neither strong nor conclusive in any of the cases considered. Moreover, UG models exhibit appealing theoretical features, such as the possibility of explaining both late time accelerated expansion and the inflationary epoch as a single gravitational phenomenon, a unification that is not achievable within the standard cosmological model with a fixed cosmological constant. These features provide compelling motivation for the continued investigation of this class of models.

\section{Acknowledgments}
The authors would like to thank S. J. Landau for their helpful comments.


The authors are supported by CONICET Grant No. PIP 11220200100729CO, and Grants No.SG002 from UNLP.

\newpage
\bibliographystyle{apsrev}
\bibliography{testmog}
\end{document}